# Exciton radiative lifetime in a monoatomic carbon chain

Stella Kutrovskaya,[1,2,3] Sevak Demirchyan,[1,2] Anton Osipov,[3,4] Stepan Baryshev,[5] Anton Zasedatelev,[5,6] Pavlos Lagoudakis,[5,6] and Alexey Kavokin[1,2,6,7,8]

[1] School of Science, Westlake University, 18 Shilongshan Road, Hangzhou 310024, Zhejiang Province, China
[2] Institute of Natural Sciences, Westlake Institute for Advanced Study, 18 Shilongshan Road, Hangzhou 310024, Zhejiang Province, China
[3] Department of Physics and Applied Mathematics, Stoletov Vladimir State University, 600000 Gorkii street, Vladimir, Russia
[4] ILIT RAS | Branch of FSRC \Crystallography and Photonics" RAS,1 Svyatoozerskaya, Shatura, 140700, Moscow region, Russia
[5] Skolkovo Institute of Science and Technology, 30 Bolshoy Boulevard, bld. 1, 121205 Moscow, Russia
[6] Physics and Astronomy, University of Southampton, Highfield, Southampton, SO171BJ, United Kingdom
[7] Russian Quantum Center, Skolkovo IC, Bolshoy Bulvar 30, bld. 1, Moscow 121205, Russia
[8] NTI Center for Quantum Communications, National University of Science and Technology MISiS, Moscow 119049, Russia

Linear carbon-based materials such as polyyne and cumulene oligomers provide a versatile platform for nano-physics and engineering. Direct gap quasi-1D polyyne structures are promising for the observation of strong and unusual excitonic effects arising due to the two-dimensional quantum confinement. Recently, we reported on the observation of sharp exciton peaks in low temperature photoluminescence spectra of polyyne chains [1]. Here, we analyse the time-resolved optical response of this system. We extend the non-local dielectric response theory to predict the exciton radiative lifetime dependence on the bandgap value and on the length of the chain. A good agreement between the experiment and the theory is achieved.

**Introduction.** The hybridization of s– and p–orbitals of carbon carbon atoms gives rise to the formation of a rich variety of allotropic forms of carbon crystals including $sp^1$ (carbyne), $sp^2$ (graphite), $sp^{2+}$ (fullerenes), $sp^3$ (diamond) etc. For a long time, only diamond, graphite and coal were widely known.

Fullerenes have been discovered in 1985 [2, 3], then carbon nanotubes [4] and graphene [5] entered into play.

Despite of the rapid progress in the methods of synthesis of low-dimensional carbon based structures, the controllable synthesis of linear $sp^1$-hybridized carbon materials is still quite challenging. The interest in $sp^1$-carbon is flaring up since 1969, when chaoite in shock-fused graphite gneiss from the Ries crater in Bavaria was discovered [6]. Long linear carbon chains (LLCC) exist in two allotropic forms, namely, polyyne characterized by alternating single and triple electronic bonds between atoms and cumulene characterised by double bonds. It has been shown in multiple publications that polyyne represents a direct bandgap semiconductor, while infinite cumulene chains are expected to be metallic [7, 8]. Due to its direct gap belonging to the optical frequency range, polyyne attracted much interest from the point of view of optical effects and applications. In particular, a possibility to tune the bandgap of polyyne across nearly the full visible range by changing the length of the chain offers an opportunity to use LLCCs as nanoscale light sources [9, 10]. In order to achieve a further in-site into quantum and optical properties of polyyne-based nanostructures, we studied recently the excitonic spectra that dominate the photoluminescence of polyyne chains at low temperatures [1]. The recent observation of excitons and trions in LLCCs has a fundamental importance as it presents the first experimental evidence of bright exciton states in monoatomic chains.

It crowns an effort of many groups who provided the excitonic gaps calculation in one-dimensional carbon chains [11, 12] and predicted the excitonic features in carbon chains [13]. This study has revealed the radiative lifetime of an exciton of the order of 1ns [1]. In addition, it was found that the increase of the bandgap width with the decrease of the LLCC length is accompanied by a significant shortening of the exciton radiative lifetime.

The present work is aimed at the interpretation of this result and modelling of radiative properties of exciton states in linear carbon chains of finite lengths. We have extended the non-local dielectric response theory formulated for excitons in semiconductor quantum wells, wires and dots in 1990s [14] in order to account for the specific features of excitons in finite-size monoatomic chains. Comparison of the predictions of the model with the data allowed to extract a precious information on the main excitonic characteristics such as the oscillator strength, Bohr radius and binding energy.

**The model.** We shall theoretically describe the radiative decay of excitons in kinked monoatomic carbon chains confined between gold nanoparticles (NPs). First, we neglect the kink effect and assume that an exciton as a whole particle is confined between two gold NPs. Since the distance between NPs is comparable with the wavelength of light, each chain may be conveniently modelled as a large and strongly anisotropic quantum dot (QD) (Fig. 1).

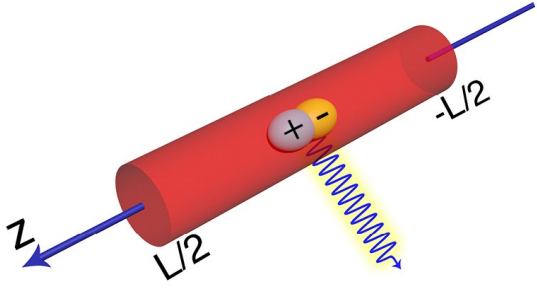

FIG. 1: Schematically depicts an exciton confined in an elongated cylindrical quantum dot that models a carbon chain sandwiched between two gold NPs.

Following the non-local dielectric response model developed in Ref. [14], the exciton radiative life-time in an individual QD can be found as:

$$\tau_{QD} = \frac{1}{2\Gamma_0^{QD}}, \quad (1)$$

where

$$\Gamma_0^{QD} = \frac{1}{6}\omega_{LT} k_0^3 a_B^3 \left[\int \cos(\vec{k}\vec{r}) \Phi(\vec{r}) d\vec{r}\right]^2. \quad (2)$$

Here $\omega_{LT}$ is the exciton longitudinal-transverse splitting in a bulk crystal, $k^2 = \epsilon_b k_0^2$, $k_0 = \omega_0/c$, $\epsilon_b$ is the background dielectric constant, $\omega_0$ is the exciton resonance frequency, $a_B$ is the three-dimensional exciton Bohr-radius, $\Phi(\vec{r}) = \Psi(\vec{r}_e, \vec{r}_h)$ is the full exciton wave function taken with equal electron and hole coordinates.

The QD is strongly elongated, and the exciton center of mass wavefunction is a one-dimensional function quantized due to the exciton confinement between the ends of the chain. We shall use the trial wave function of electron-hole relative motion along the axis of the chain introduced in Ref. [15] and assume that the center of mass wave function may be represented in a form:

$$\Psi(\vec{r}_e, \vec{r}_h) = U_h(\vec{\rho}_h) U_e(\vec{\rho}_e) f(z) F(Z) = \frac{\sqrt{a}}{\pi R_h R_e} \sqrt{\frac{2}{L}} \times$$
$$\times \exp\left(-\frac{\rho_h^2}{2R_h^2}\right) \exp\left(-\frac{\rho_e^2}{2R_e^2}\right) \exp(-a|z|) \cos\left(\frac{\pi}{L} Z\right), \quad (3)$$

where $z = z_e - z_h$, $Z = (z_e m_e + z_h m_h)/(m_e + m_h)$ is the center of mass coordinate, $R_h = 0.1 nm$, $R_e = 0.25 nm$, $a = 0.351 nm^{-1}$.

The exciton longitudinal-transverse splitting can be expressed as [16]:

$$w_{LT} = \frac{4e^2 |p_{cv}|^2}{\hbar \omega_0^2 m_0^2 \epsilon_b a_B^3}, \quad (4)$$

where $m_0$ is the free electron mass, $|p_{cv}|^2$ is the interband dipole matrix element. Combining equations (2)-(4) we obtain:

$$\Gamma_0^{QD} = \frac{2}{3} \frac{e^2 |p_{cv}|^2 \omega_0}{\hbar m_0^2 c^3 \epsilon_B} \frac{2a}{L} \left[\int_0^\infty 2\pi\rho U_e(\rho) U_h(\rho) d\rho\right]^2$$
$$\times \left[\int_{-L/2}^{L/2} \cos(kZ) \cos\left(\frac{\pi}{L} Z\right) dZ\right]^2. \quad (5)$$

Using the k p - method of band-structure calculations we estimate the interband matrix element as [17]:

$$|p_{cv}|^2 = \left(\frac{m_0^2}{m_c} - m_0\right) \frac{E_g}{2}, \quad (6)$$

where $m_c = 0.078 m_0$ is the electron effective mass in the conduction band [1]. Thus, the exciton radiative lifetime can be found as:

$$\tau_{QD} = \frac{3\hbar^2 c^3 \epsilon_B}{2e^2 (1/m_c - 1/m_0) E_g^2} \frac{L}{2a} I_{eh}^{-2} [J(L)]^{-2}, \quad (7)$$

where

$$I_{eh} = \int_0^\infty 2\pi\rho U_e(\rho) U_h(\rho) d\rho, \quad (8)$$

and

$$J(L) = \frac{\frac{2\pi}{L} \cos\left(\frac{kL}{2}\right)}{\frac{\pi^2}{L^2} - k^2}. \quad (9)$$

**Results and discussion.** Figure 2 shows the experimental low temperature time-resolved photoluminescence spectra (TRPL) of linear carbon chains adapted from Ref. [1]. TRPL curves of different color correspond to different spectral bands, as shown in the inset. Each specific spectral band is attributed to a fixed number of atoms contained in a straight part of the chain. We note that virtually all the studied chains contain kinks diving them into straight parts containing even numbers of carbon atoms. The band gap width in polyyne depends on a number of atoms in the straight part of the chain confined between the neighboring kinks or between the first (last) kink and the gold NP. In our experiments, the gap width varies from 2.2 to 2.8eV for the straight parts of chains ranging in length form 10 to 20 atoms. In order to analyze this set of data, first, we shall fix the average value of the band gap for whole chain taking $E_g = \hbar\omega_0 = 2.5 eV$ and study the dependence of the exciton radiative lifetime on the chain length. Our calculation shows that excitons in the polyyne chains of $20 - 100 nm$ lengths are characterized by the radiative lifetimes of the order of 1ns, which nicely agrees with the experiment data. Having obtained, within the framework of the developed model, the correct order of magnitude for the radiative lifetime for the experimentally observed lengths, our next step is to study the dependence of the radiative lifetime on the bandgap width. To do this, we will fix the chain length and plot the calculated exciton lifetime as a function of the bandgap value, which depends on the number of atoms in the linear parts of polyyne chain between kinks in our experiments. Fig. 3 presents a dependence of the exciton radiative lifetime on the bandgap for the straight parts of the chains of different lengths. The results of this calculation are compared to the experimental data extracted from Fig. 2 and shown by stars in Fig. 3.

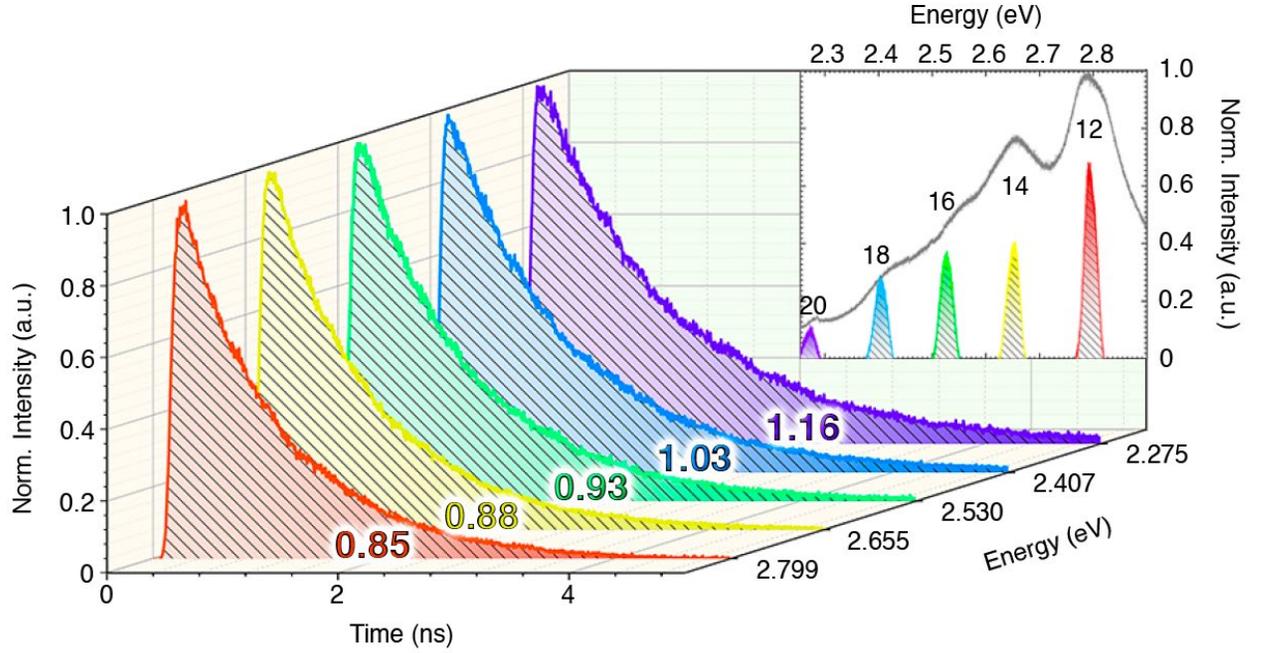

FIG. 2: The experimentally measured time resolved photoluminescence (TRPL) spectra of gold-stabilized linear carbon chains acquired at 4K. The inset shows the spectral bands corresponding to each TRPL measurement. The selected bands correspond to the excitonic transitions in straight chains of a different within the 5*nm* accuracy (the number of atoms in the chain is indicated on the top of the corresponding spectral resonance). The extracted exciton radiative lifetime (in nanoseconds) is indicated for each band.

It is important to note that the experimentally studied chains are of an average total length (the spacing between two gold NPs) of about 50nm. On the other hand, the length of the straight parts of the chains separated by kinks is in the range of 1 − 3nm. This estimate is obtained having in mind that we usually have 10 − 20 atoms between the neighboring kinks [1] and the interatomic distances in polyyne are known to be 0.133nm (C–C) and 0.123nm (C≡C), respectively. The length of the straight parts (number of atoms) determines the transition frequency [1]. Clearly, each individual chain may be composed of straight parts of different lengths characterized by different energy gaps. We assume that the exciton as a whole may resonantly transfer between identical straight parts of different chains in a bundle.

This is a reasonable assumption as the bandgap energies in the straight parts of the chains of the same length match, and the exciton transfer between them would not cost any energy. Based on this assumption, we consider the distance between two gold NPs as a characteristic size of the nanoobject that confines excitons contributing to the radiative decay signal Fig. 4. This assumption is confirmed by two experimental observations. Ref. [1] demonstrated the thermal dissociation of an exciton as a result of hopping of one of the carriers, which indicates that hops between parallel chains are, in principle, possible, and nothing prevents the exciton as a whole from hopping onto on of the neighboring chains provided that it finds a section characterised by the same energy as one where it departs from. The second argument is that if the exciton would be strictly confined to every individual straight part of a chain between two neighboring kinks, than both the dependence of the lifetime on the bandgap and the absolute value of the lifetime would be strongly different. It was shown in [18, 19] that the lifetime of an exciton critically depends on the relationship between the size of the exciton, the size of the QD, and the wavelength at the exciton resonance. Clearly, in the case where the size of the quantum dot is significantly smaller than the wavelength of light, the retardation effects can be neglected, but in the case where the size of the quantum dot approaches the wavelength of light, they begin to play an essential role and cause a strong increase of the radiative lifetime.

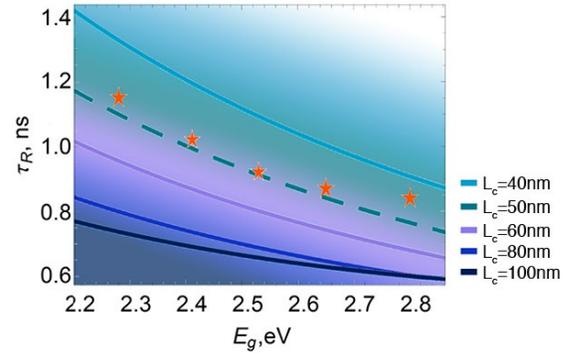

FIG. 3: The radiative decay time of excitons versus the band gap energy for different fixed distances between NPs $L_c = 40nm; 60nm; 80nm; 100nm$ respectively. Stars display the experimental data taken from Figure 2. The best fit is achieved at $L_c = 50nm$ (shown by the dotted line).

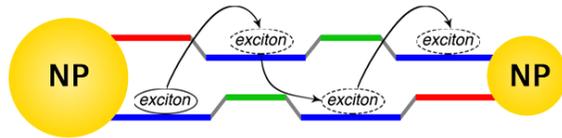

FIG. 4: A schematic that illustrates the confinement of an exciton in a bundle of polyyne chains sandwiched between two gold NPs and separated into straight parts by the kinks. As all excitons belonging to the straight parts of the same length emit at the same energy, in a bundle that contains a sufficiently large number of chains, the volume occupied by excitons that contribute to the radiative recombination at a given energy is limited by the distance between the gold NPs.

In conclusion, we calculated the exciton radiative lifetime in bundles of monoatomic linear carbon chains sandwiched between gold NPs in the framework of the non-local dielectric response theory. We have found a good agreement between the results of the present model and the data of low temperature time-resolved photoluminescence measurements. The time-resolved optical response of excitons in carbon chains is found to be dependent on the bandgap of the chain and the lengths of straight parts of the chains that provides a control tool for fine-tuning of the radiative properties of carbon chains for applications in carbon lasers and light-emitting diodes.

**Acknowledgement.** The work is supported by the Westlake University, project 041020100118 and the Program 2018R01002 funded by Leading Innovative and Entrepreneur Team Introduction Program of Zhejiang, by MSHE within the State assignment VlSU 0635-2020-0013. SD acknowledges the support from the Grant of the President of the Russian Federation for state support of young Russian scientists No. -5318.2021.1.2. AK acknowledges the support from Rosatom within the Road map for quantum computing.

[1] Kutrovskaya S., Osipov A., Baryshev S., Zasedatelev A., Samyshkin V., Demirchyan S., Pulci O., Grassano D., Gontrani L., Hartmann R.R., Portnoi M.E., Kucherik A., Lagoudakis P.G., Kavokin A. Excitonic Fine Structure in Emission of Linear Carbon Chains. Nano Letters, 20(9), 6502-6509 (2020).

[2] Kroto H. W., Heath J. R., O'Brien S. C., Curl R.F., Smalley R. E. C60: Buckminsterfullerene. Nature, 318(6042), 162-163 (1985).

[3] Haddon R. C., Brus L. E., Raghavachari K. Electronic structure and bonding in icosahedral C60. Chemical Physics Letters, 125(5-6), 459-464 (1986).

[4] Iijima S., Ichihashi T. Single-Shell Carbon Nanotubes of 1-nm Diameter. Nature, 363, 603{605 (1993).

[5] Novoselov K., Geim A., Morozov S., Jiang D., Zhang Y., Dubonos S., Grigorieva I., Firsov A. Electric Field E ect in Atomically Thin Carbon Films. Science, 306, 666{669 (2004).

[6] El Goresy A., Donnay G. A new allotropic form of carbon from the Ries crater. Science, 161(3839), 363-364 (1968).

[7] Casari C. S., Tommasini M., Tykwinski R. R., Milani A. Carbon-atom wires: 1-D systems with tunable proper- ties. Nanoscale, 8(8), 4414-4435 (2016).

[8] Artyukhov V. I., Liu M., Yakobson B. I. Mechanically induced metal-insulator transition in carbyne. Nano letters, 14(8), 4224-4229 (2014).

[9] Cataldo F. (Ed.) Polyynes: synthesis, properties, and applications. CRC Press, (2005).

[10] Cretu O., Botello-Mendez A. R., Janowska I., Pham-Huu C., Charlier J. C., Banhart F. Electrical transport measured in atomic carbon chains. Nano letters, 13(8), 3487-3493 (2013).

[11] Mostaani E., Monserrat B., Drummond N.D., Lambert C.J. Quasiparticle and excitonic gaps of one-dimensional carbon chains. Phys. Chem. Chem. Phys. 18, 14810- 14821 (2016).

[12] Deretzis I., La Magna A. Coherent electron transport in quasi one-dimensional carbon-based systems. The European Physical Journal B, 81(1), 15 (2011).

[13] Bonabi F., Brun S. J., Pedersen T. G. Excitonic optical response of carbon chains con ned in single-walled carbon nanotubes. Physical Review B, 96(15), 155419 (2017).

[14] Ivchenko E. L., Kavokin A. V., Kochereshko V. P., Kop'ev P. S., Ledentsov N. N. Exciton resonance re ecxtion from quantum well, quantum wire and quantum dot structures. Superlattices and microstructures, 12(3), 317- 320 (1992).

[15] Kutrovskaya S., Demirchyan S., Osipov A., Baryshev S., Zasedatelev A., Lagoudakis P., Kavokin A., Exci- ton energy spectra in polyyne chains, arXiv preprint, arXiv:2007.10481 (2020).

[16] Ivchenko E. L. Excitonic polaritons in periodic quantum-well structures. Soviet physics. Solid state, 33(8), 1344-1346 (1991).

[17] Yu P. Y., Cardona M. Fundamentals of Semiconductors. Graduate Texts in Physics. Springer-Verlag Berlin Heidelberg (2010). [18] Rashba E.I., Gurgenishvili G. E. On the theory of edge absorption in semiconductors. Sov. Phys. Solid State, 4, 759-760 (1962).

[19] Gil B., Kavokin A. V. Giant exciton-light coupling in ZnO quantum dots. Applied Physics Letters, 81(4), 748- 750 (2002).